\documentclass[sn-mathphys-num]{sn-jnl}

\usepackage{graphicx}%
\usepackage{multirow}%
\usepackage{amsmath,amssymb,amsfonts}%
\usepackage{amsthm}%
\usepackage{mathrsfs}%
\usepackage[title]{appendix}%
\usepackage{xcolor}%
\usepackage{textcomp}%
\usepackage{manyfoot}%
\usepackage{booktabs}%
\usepackage{algorithm}%
\usepackage{algpseudocode}%
\usepackage{listings}%

\theoremstyle{thmstyleone}%
\theoremstyle{thmstyletwo}%
\newtheorem{remark}{Remark}%

\theoremstyle{thmstylethree}%

\begin{document}

\title[NN aided QST with constrained measurements]{Neural networks for quantum state tomography with constrained measurements}


\author[1,2]{Hailan Ma}\email{hailanma0413@gmail.com}

\author*[1]{Daoyi Dong}\email{daoyi.dong@anu.edu.au}

\author[1]{Ian R. Petersen}\email{i.r.petersen@gmail.com}

\author[3]{Chang-Jiang Huang}\email{hcj00@mail.ustc.edu.cn}

\author[3]{Guo-Yong Xiang}\email{gyxiang@ustc.edu.cn}

\affil*[1]{CIICADA Lab, School of  Engineering, The Australian National University, Canberra, ACT 2601, Australia}

\affil[2]{School of Engineering and Technology, University of New South Wales, Canberra, ACT 2600, Australia}

\affil[3]{Key Laboratory of Quantum Information, University of Science and Technology of China, Chinese Academy of Sciences, Hefei 230026, China}


\abstract{Quantum state tomography (QST) aiming at reconstructing the density matrix of a quantum state plays an important role in various emerging quantum technologies. Recognizing the challenges posed by imperfect measurement data, we develop a unified neural network(NN)-based approach for QST under constrained measurement scenarios, including limited measurement copies, incomplete measurements, and noisy measurements. Through comprehensive comparison with other estimation methods, we demonstrate that our method improves the estimation accuracy in scenarios with limited measurement resources, showcasing notable robustness in noisy measurement settings. These findings highlight the capability of NNs to enhance QST with constrained measurements.}


\maketitle

\section{Introduction}\label{sec1}

Quantum state tomography (QST) plays a significant role in verifying and benchmarking quantum tasks, including quantum computation~\cite{nielsen2010quantum}, quantum control ~\cite{dong2010quantum} and quantum communication  ~\cite{gisin2007quantum}. It involves reconstructing the state of a quantum system via quantum measurements ~\cite{dong2022quantum}. To achieve this, least-squares inversion was proposed to solve the inverse of linear equations that relate the measured quantities to the density-matrix elements of a quantum system~\cite{opatrny1997least}. Bayesian tomography constructs a state using an integral averaging over all possible quantum states with proper
weights~\cite{huszar2012adaptive,blume2010optimal}. Maximum likelihood estimation (MLE) chooses the state estimate that maximizes the probability of the observed data, whose solution usually involves many nonlinear equations~\cite{vrehavcek2001iterative,jevzek2003quantum}.  
Linear regression estimation (LRE) solves the quantum state estimation problem using a linear model and is usually combined with physical projection techniques to avoid the non-physical quantum states ~\cite{qi2013quantum}. 


The exponential scaling of parameters in a quantum state requires an exponentially increasing number of complete measurements~\cite{Measurement2001}, each of which requires a sufficient number of identical copies, posing a great challenge in practical applications. For example, measuring quantum states can be time-consuming and requires complex and costly experimental setups, which might limit the available measurements~\cite{chantasri2019quantum}. Within coherence time, one may obtain a complete but inaccurate data vector (a few copies assigned to each measurement operator) or an accurate but incomplete data vector (sufficient copies assigned to partial measurements that are experimentally easy to generate).  Another persistent challenge is systematic errors in the inherent susceptibility of quantum measurements to noise, which increases the difficulty of fully characterizing quantum states~\cite{smolin2012efficient}. These imperfect factors are collectively referred to as constrained measurements in this paper, which hinder the precise reconstruction of quantum states in practical applications.

Recently, machine learning (ML) has been utilized to address various quantum tasks~\cite{carleo2017solving,lennon2019efficiently,fosel2018reinforcement,huang2020realization}. ML's capability to handle complex data makes it suitable for extracting useful patterns from imperfect measured data for QST issues~\cite{biamonte2017quantum}. For example, neural networks (NNs) act as a useful tool to denoise the state-preparation-and-measurement errors~\cite{palmieri2020experimental}. Conditional generative adversarial networks have been introduced for the reconstruction of quantum optical states~\cite{ahmed2021quantum,ahmed2021classification} and convolutional neural networks (CNNs) have been applied to estimate quantum states from incomplete measurements~\cite{lohani2020machine,danaci2021machine,lohani2021experimental}. Despite these advancements, the exploration of how NNs enhance QST with various constrained measurements, especially the advantage over traditional methods has been largely unexplored.

In this context, we develop an NN-based QST method designed for the effective reconstruction of quantum states under measurement constraints. In particular, we design a straightforward yet expressive NN architecture composed of multiple layers of fully connected networks. These networks  map measured statistics to intermediate real vectors which are then used to construct lower-triangular matrices, ultimately leading to the generation of density matrices. In contrast to other methods that rely on prior assumptions (e.g., the low-rank assumption in compressed sensing~\cite{gross2010quantum}), our approach offers a unified solution applicable to a variety of constrained measurement scenarios common in quantum applications. To benchmark the performance, we compare our method with MLE, LRE, and a CNN-based QST method~\cite{lohani2020machine}. An improvement in average fidelity demonstrates the advantage of our approach in reconstructing quantum states with limited measurement resources. Moreover, our method has demonstrated favorable robustness in estimating quantum states when tomographic measurements suffer from noise. These findings highlight the potential of NNs to enhance QST beyond the traditional dependence on extensive measurements.

The rest of this paper is organized as follows. In Section \ref{sec:method}, the DNN-QST method is presented in detail. Numerical results of QST with limited resources, i.e., few measurement copies and incomplete measurements are provided in Section \ref{sec:limitedresources}.  Section \ref{sec:noise} investigates the performance of QST with noise. Concluding remarks are given in Section \ref{sec:conclusion}.

	\section{Deep neural networks based quantum state tomography}\label{sec:method}

In this section, a machine learning-aided quantum state tomography method (called DNN-QST) that takes advantage of deep neural networks to reconstruct quantum states is proposed. Quantum state generation and its representation are first presented, and then quantum measurement settings are introduced. Finally, the DNN-QST method is provided in detail.

\subsection{Quantum state generation and its representation}
In this work, we first consider pure states.  Let $\mathbb{U}^d$ be the set of all $d$-dimensional unitary operators, i.e., any $U \in \mathbb{U}^d$ satisfies $UU^{\dagger} =U^{\dagger}U = \mathbb{I}_d$. Owing to its invariance under group multiplication (i.e., any region of $\mathbb{U}^d$ carries the same weight in a group average)~\cite{mezzadri2006generate}, the Haar metric is utilized as a probability measure on a compact group for generating random unitary matrix. As such, random pure states can be generated by performing random unitary transformations following the Haar measure~\cite{qi2017adaptive,danaci2021machine}, which can be formulated as   $|\psi\rangle_{haar} = U_{haar}  |\psi_0\rangle$, where $|\psi_0\rangle$ is a fixed pure state. Then, we consider mixed states in the following form
\begin{equation}
	\rho_p = p |\psi\rangle \langle \psi|_{haar} +(1-p) \frac{\mathbb{I}_d}{d},
	\label{eq:mixed states}
\end{equation}
where $p\in(0,1)$ represents the ratio of pure element. 
The purity of the quantum state in Eq. (\ref{eq:mixed states}) is closely related to the value of $p$, namely, $\textup{Tr}(\rho_p^2)= p^2 (1 - 1/d) + 1/d$.
Let $\mathcal{N}(0,1)_d$ represent random normal distributions of size $d \times d$ with zero mean and unity variance. Then, we consider random mixed states generated from the Ginibre ensembles using the Hilbert-Schmidt metric~\cite{al2010random}, given  by
\begin{equation}
	\rho_{mixed}=\frac{\rho_G \rho_G^{\dagger}}{\textup{Tr}(\rho_G \rho_G^{\dagger})},\quad  \rho_G=\mathcal{N}(0,1)_d+\textup{i}\mathcal{N}(0,1)_d.
	\label{eq:ginibre state}
\end{equation}

When only considering pure states, splitting $\psi$ into real and imaginary parts is  natural solution for parameterizing quantum states. Here, we try to propose a general QST method that can be applied to mixed states. To achieve a unified notation for pure states and mixed states, a density matrix $\rho$, i.e., a $d \times d$ matrix is utilized to describe a quantum state. 
Although there are many ways to generate a Hermitian operator, the condition of positivity is difficult to guarantee. Given any lower triangular matrix $\rho_L$, a physical density matrix can be obtained as
\begin{equation}
	\rho = \frac{\rho_L\rho_L^{\dagger}}{\textup{Tr}(\rho_L\rho_L^{\dagger})}.
	\label{eq:tau2rho}
\end{equation}
$\rho$ in Eq. (\ref{eq:tau2rho}) satisfies the three conditions of Hermitian, positive semidefinite, and unit trace.  
In addition, according to the Cholesky decomposition~\cite{higham1990analysis}, for any density matrix $\rho$, there exists a lower triangular matrix $\rho_L$ that achieves $
\rho_L\rho_L^{\dagger} = \rho$. Note that a tiny perturbation term (e.g., $\epsilon=10^{-7}$) is usually added to the simulated pure states to avoid convergence  issues using the Cholesky decomposition~\cite{higham1990analysis}
\begin{equation}
	\rho=(1-\epsilon)|\psi\rangle\langle\psi|+\frac{\epsilon}{d} \mathbb{I}_d.
\end{equation}

As such, any density matrix $\rho$ can be parameterized as a corresponding lower triangular matrix $\rho_L$, which has been widely used in previous work~\cite{ma2021neural,ahmed2021quantum,ahmed2021classification,wang2022ultrafast}.  By comparsion, the Bloch representation requires additional efforts to pull a non-physical state into a physical state~\cite{koutny2022neural}. Hence, the search for a physical quantum state can be converted to the search for a lower triangular matrix, which can be further transformed into a real vector by splitting the real and imaginary parts.

\subsection{Quantum measurement settings}\label{sub:measurement}

\begin{table*}[ht]
	\caption{The measurement bases used in cube and MUB~\cite{adamson2010improving}.}
	\label{Table:cube-mub}
	\begin{tabular*}{\textwidth}{@{}l*{15}{@{\extracolsep{0pt plus12pt}}l}}
			\hline 
			cube &MUB\\
			\hline  
			$|HH\rangle$,$|HV\rangle$,$|VH\rangle$,$|VV\rangle$
			&$|HH\rangle$,$|HV\rangle$,$|VH\rangle$,$|VV\rangle$\\
			$|HD\rangle$,$|HA\rangle$,$|VD\rangle$,$|VA\rangle$ & $|RD\rangle$,$|RA\rangle$,$|LD\rangle$,$|LA\rangle$\\
			$|HR\rangle$,$|HL\rangle$,$|VR\rangle$,$|VL\rangle$ & $|DR\rangle$,$|DL\rangle$,$|AR\rangle$,$|AL\rangle$\\
			$|DH\rangle$,$|DV\rangle$,$|AH\rangle$,$|AV\rangle$ & $\frac{1}{\sqrt{2}}(|RL\rangle+{\rm{i}}|LR\rangle)$, $\frac{1}{\sqrt{2}}(|RL\rangle-{\rm{i}}|LR\rangle)$,\\
			$|DD\rangle$,$|DA\rangle$,$|AD\rangle$,$|AA\rangle$ &  $\frac{1}{\sqrt{2}}(|RR\rangle+{\rm{i}}|LL\rangle)$, $\frac{1}{\sqrt{2}}(|RR\rangle-{\rm{i}}|LL\rangle)$\\
			$|DR\rangle$,$|DL\rangle$,$|AR\rangle$,$|AL\rangle$ & $\frac{1}{\sqrt{2}}(|RV\rangle+{\rm{i}}|LH\rangle)$, $\frac{1}{\sqrt{2}}(|RV\rangle-{\rm{i}}|LH\rangle)$,\\
			$|RH\rangle$,$|RV\rangle$,$|LH\rangle$,$|LV\rangle$ & $\frac{1}{\sqrt{2}}(|RH\rangle+{\rm{i}}|LV\rangle)$, $\frac{1}{\sqrt{2}}(|RH\rangle-{\rm{i}}|LV\rangle)$\\
			$|RD\rangle$,$|RA\rangle$,$|LD\rangle$,$|LA\rangle$ & \\
			$|RR\rangle$,$|RL\rangle$,$|LR\rangle$,$|LL\rangle$ &\\
			\hline
		\end{tabular*}
	\end{table*}
	
	Theoretically, the minimum number of measurements to fully characterize a quantum state with dimension $d$ is $d^2$~\cite{Measurement2001}. For example, a linearly independent set of projectors from pairwise combinations of eigenstates of the Pauli operators can be constructed~\cite{Measurement2001}.  An improved estimate of 2-qubit quantum states can be obtained when tomography is performed using projections onto 36 tensor products of Pauli eigenstates~\cite{de2008choice}. In addition, mutually unbiased bases where all inner products between projectors of different bases are equal to $\frac{1}{d}$ have the potential to maximize information extraction per measurement ~\cite{adamson2010improving}. Hence, the measurement operators play an important role in the estimation problem. 
	
	Let the eigenvectors of $\sigma_x$ be the diagonal state $|D\rangle$ and the anti-diagonal state $|A\rangle$. Define the eigenvectors of $\sigma_y$ as the left circular polarization state $|L\rangle$ and the right circular polarization state $|R\rangle$ and take the eigenvectors of $\sigma_z$ as the horizontal state $|H\rangle$ and the vertical state $|V\rangle$.
	These six basis states form the 1-qubit Pauli measurement~\cite{de2008choice}, $	P_{pauli}=\{|H\rangle,|V\rangle,|D\rangle,|A\rangle,|R\rangle,|L\rangle\}$. They are informationally complete for reconstructing 1-qubit quantum states. 
	
	In this work, we consider two types of measurement settings: (i) tensor products of Pauli matrices, which is also called the cube measurement~\cite{de2008choice}, (ii) the mutually unbiased bases (MUB) measurement~\cite{adamson2010improving}. For $n$-qubits, there are $6^n$ measurement basis operators involved in cube measurement and $(4^n+2^n)$ involved in MUB measurement. The measurement basis states for cube and MUB are usually grouped into sets, each containing $d=2^n$ orthogonal projectors.  Typically, the $6^n$ measurement operators in the cube measurement are arranged into  $3^n$  sets. Similarly, the  MUB measurements are grouped into  $(2^n+1)$ sets.  For the 2-qubit case, their measurement basis states are summarized in Table~\ref{Table:cube-mub}. Denoting each ideal measurement operator as $M_i$, the presence of noise may lead to the utilization of perturbed operators $\tilde{M}_i$. This disparity between $M_i$ and $\tilde{M}_i$ consequently impedes the reconstruction process.

	\subsection{DNN-QST}
	
	\begin{figure}[h]
		\centering
		\includegraphics[width=0.7\textwidth]{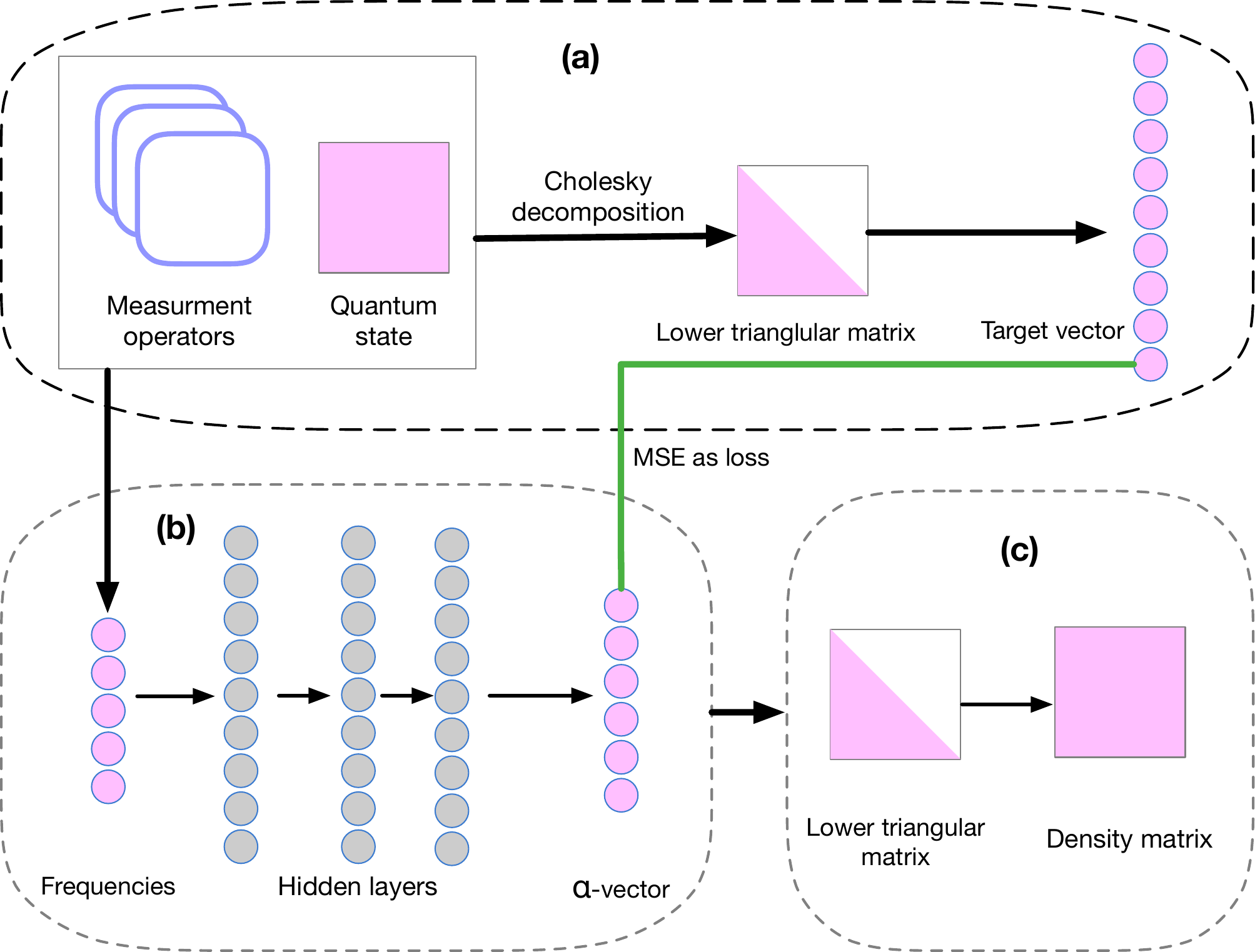}
		\caption{Schematic of the DNN-QST approach. (a) Obtain the measured data and compute the Cholesky decomposition of the density matrices as the target vectors; (b) A multi-layer NN maps the frequencies to the $\bold{\alpha}$-vector; (c) Obtain the quantum states from the networks' output.}
		\label{fig:framework}
	\end{figure}
	
According to the universal approximation theorem ~\cite{lecun2015deep},  any continuous function on a compact subset of $\mathbb{R}^n$ can be theoretically approximated by deep NNss with multiple layers of fully connected neural networks. Hence, multi-layer NNs can be constructed to approximate a function mapping from measured frequencies to physical density matrices. As such, QST is transformed into a regression problem, which can be solved in a supervised learning procedure. The input can be the observed frequencies, which depend on the choice of
	measurement operators. Recall the measurements are usually grouped into different sets (see Subsection~\ref{sub:measurement} for detailed information), the dimension of the input vector is determined by the number of sets among the measurements. In particular, the dimension of the input feature can be calculated as $dK$, with $K$ the number of measurement sets. In that case, the input vector can be denoted as $\mathbf{f}=[f_1,f_2,...f_{dK}]\in \mathbb{R}^{dK}$. For example, for the complete measurement case, its input would be $6^n$-dimensional feature vectors for the cube measurement and  $2^n(2^n+1)$-dimensional feature vectors for the MUB measurement, respectively. For the incomplete case, the input dimension would decay. In general, any measurement setting that can be grouped according to certain rules can be applied to our method. For example, one can establish a measurement setting using random square root measurements~\cite{motka2017efficient}.

	Recall a lower triangular matrix $\rho_L$ is correlated with a positive Hermitian matrix $\rho$ in Eq. (\ref{eq:tau2rho}). When designing deep NNs to reconstruct a quantum state, a lower triangular matrix $\rho_L$ can be used as an intermediate state.  $\rho_L$ is a complex matrix with real and positive diagonal entries and complex off-diagonal elements and thus can be further represented as a real vector with length $d^2$ (defined as the $\alpha$-vector in this work). In particular, the positive diagonal entries are first ordered, and then the real and imaginary parts of off-diagonal elements are considered.  As such,  the output layer can be designed with $d^2$ neurons to represent the $\alpha$-vector, which can be utilized to produce physical density matrices with additional transformations.

	The schematic of DNN-QST is summarized in Fig.~\ref{fig:framework}. In (a), samples are created through the generation of different density matrices, whose target vectors are calculated by the Cholesky decomposition. Then measured frequencies are gathered through simuations of the measurement process using density matrices and measurement operators. Denote the training sample as $\{\textbf{f},\alpha \}$, where $\textbf{f}\in \mathbb{R}^{dK}$ represents the measured frequencies on quantum state $\rho$ and $\alpha\in \mathbb{R}^{d^2}$  represents the real vector obtained by performing Cholesky decomposition on $\rho$ (equivalent to the ground truth of $\rho$). In (b),  an architecture that includes input-hidden-output layers is utilized as a parameterized function to map a feature vector $\textbf{f}$ comprising of the measurement outcomes to the $\alpha$-vector. The loss function for training the NNs adopts the mean squared error (MSE) between the reconstructed $\alpha$-vector (from the output of the NNs) and the expected target vector (from the decomposition of the density matrix). Then, the networks are trained using a gradient descent algorithm with AdamOptimizer to minimize the MSE~\cite{kingma2014adam}. 
	
	The detailed training procedures of DNN-QST are summarized in Algorithm~\ref{alg1}. Practically, the entire training dataset is divided into mini-batches to be fed into the networks one by one. An epoch elapses when an entire dataset is passed forward and backward through the NNs exactly one time (line 8-15 in Algorithm~\ref{alg1}). After the weights of the NNs have been learned,  the $\alpha$-vector is transformed into a physical density matrix in (c). Unlike the training process in (b), a procedure of $\alpha \rightarrow \rho_L \rightarrow \rho$ is required to obtain physically valid quantum states. Although the first stage requires many iterations to optimize, the second stage only requires a single feed-forward calculation without iterations. From this perspective, DNN-QST is efficient in reconstructing quantum states.

\begin{algorithm}[h]
	\caption{\small Procedures for DNN-QST.}\label{alg1}
	\begin{algorithmic}[1]
\State \textbf{Input:} Training samples with density matrices $\{\rho\}$, batch size $B$, epoch size $T$
\State \textbf{Output:} The approximated function $\mathcal{G}(\xi): \textbf{f} \rightarrow \alpha$ with weights $\xi$
		\For{each sample in the training dataset $j=1,2,...$}
		\State Perform measurements on the quantum state $\rho_j$ and collect the measurement outcomes into measured frequencies $\textbf{f}_j=[f_1,f_2,...f_{dK}]$
		\State Calculate the Cholesky decomposition on $\rho_j$ to obtain a lower-triangular matrix and a real vector $\alpha_j$
		\EndFor
		\State Randomly initialize weights of NN, i.e., $\xi$
		\For{$epoch=1,2,\cdots, T$}
		\State Randomly shuffle all the training samples
		\While{not all training samples have been visited}
		\State Sequentially select a mini-batch of $B$ samples from all the training set
		\For{each sample in the mini-batch $j=1,2,...,B$}
		\State Forward the neural networks as $\hat{\alpha}_j=\mathcal{G}(\mathbf{f}_j;\xi)$
		\EndFor
		\State Backward by minimizing MSE, i.e., $(\sum_ j |\hat{\alpha}_j- \alpha_j|^2)/B$
		\State Update $\xi$ with gradient descent
		\EndWhile
		\EndFor 
	\end{algorithmic}
\end{algorithm}

	\begin{remark}
		For the case of a few copies, we utilize ideal measurement operators $\{M_i\}$ to simulate the measurement process. Here, the number of copies are limited, and the obtained frequencies have large errors. For the case of incomplete measurements, we use sufficient copies to simulate the measurement process, but the total measurements are not informational complete. For the case of noisy measurements, we utilize the noisy operator $\{\tilde{M}_i \}$ to simulate the measurement process.  
	\end{remark}

\section{Results on QST with limited measurement resources}\label{sec:limitedresources}

In this section, parameter settings are first introduced, and then QST using few measurement copies and incomplete measurements is investigated and analyzed to demonstrate the effectiveness of our approach. 

\subsection{Parameter settings}\label{subsec:parameter}
For each case, a large number of samples are generated by randomly sampling the parameters and split for different purposes.  98,800 samples are used for training the parameters of DNN-QST. 1000 samples are used for testing the model. Hence, training data and testing data are generated from the same distributions in our work.  In addition, those 1000 samples are also tested on LRE and MLE. The measured frequencies are obtained by simulating the measurement process with a finite number of copies. In this work, we introduce $S$ to represent the average number of copies for each measurement, with the total number of copies being $SK$, where $K$ represents the number of measurements. For the case of few copies, we set $S$ to be different values, such as $\{10,20,20,40,60,80,100\}$ to investigate the performance under varying numbers of copies. For the cases of incomplete and noisy measurements, $S$ is fixed  as 100,000 to reduce the impact of limited measurement copies.

For simulations, the Pytorch framework is utilized to construct the NNs to run the model. In particular, three hidden layers are utilized, with each hidden layer using the same neurons. The number of neurons for the hidden layers is set as 128 for 2-qubit pure states and is set as 256 for 2-qubit mixed states and 3-qubit states. All neurons are activated by the leaky Relu function except the one before the $\alpha$-vector. 

We also implement a CNN-based QST method~\cite{lohani2020machine}, where the measured frequency vectors are first encoded as an image (e.g., $6\times6$ pixel for 2-qubit states with the cube measurements) and are then processed via conventional layers of kernel $2\times 2$, stride length of 1, 25 feature mapping and max-pooling layers of kernel $2\times 2$, followed by a fully connected layer.
For MLE,  the iteration process is terminated once the gap of infidelity between two successive runs falls below $10^{-8}$. For LRE, a fast version of projection is performed to project a  Hermitian operator with negative eigenvalues to a physical density operator~\cite{smolin2012efficient}. The training of  98,800 samples is conducted on GPU with 2000 epochs and 256 batch sizes for CNN and our method.  For inference time, we conduct 1,000 times of reconstructing the density matrix on CPU and report the average time per inference in Table~\ref{table:compare}. MLE and LRE do not involve a training process and their computations are both accomplished on CPU. To demonstrate the efficiency of the DNN-QST method,  the infidelity of two states is defined as $\bar{F}=1-F(\rho,\sigma)$, with the fidelity between the reconstructed state and the expected state defined as $F(\rho,\sigma)=|\textup{Tr}(\sqrt{\sqrt{\rho}\sigma\sqrt{\rho}})|$.

\begin{table*}
	\centering
	\caption{Comparison of computation efforts for four methods when reconstructing pure states using cube measurements.}
	\begin{tabular}{c|ccc|cccc}
		\toprule[2pt]
		& \multicolumn{3}{c|}{2 qubit} & \multicolumn{3}{c}{3 qubit}\\
		\midrule[2pt]
		Methods&    Params & Train & Infer&    Params & Train & Infer \\
		LRE  & N/A &  N/A  & 18.21ms & N/A  & N/A & 141.52ms\\
		MLE &  N/A & N/A & 21.96ms &  N/A  & N/A & 145.28ms \\
		CNN &  765K & 1h & 1.25ms  & 2.8M & 5h & 1.49ms\\
		DNN & 56K  & 38min & 1.26ms  & 85K &  48 min & 1.40ms \\
		\bottomrule[2pt]
	\end{tabular}
	\label{table:compare}
\end{table*}

\subsection{Few copies}


In real applications, performing one measurement on quantum systems may return a statistical value of probability, i.e., frequency. Considering that an individual outcome only provides limited information on the estimated state, an infinite number of measurements is required to determine a quantum state precisely. However, only finite measurements are available and performed on identically prepared states, with observed outcomes gathered as frequencies. A low number of measurement copies will cause the measured frequencies to deviate from the true probabilities. 

	\begin{figure}[h]
	\centering
	\includegraphics[width=1.0\textwidth]{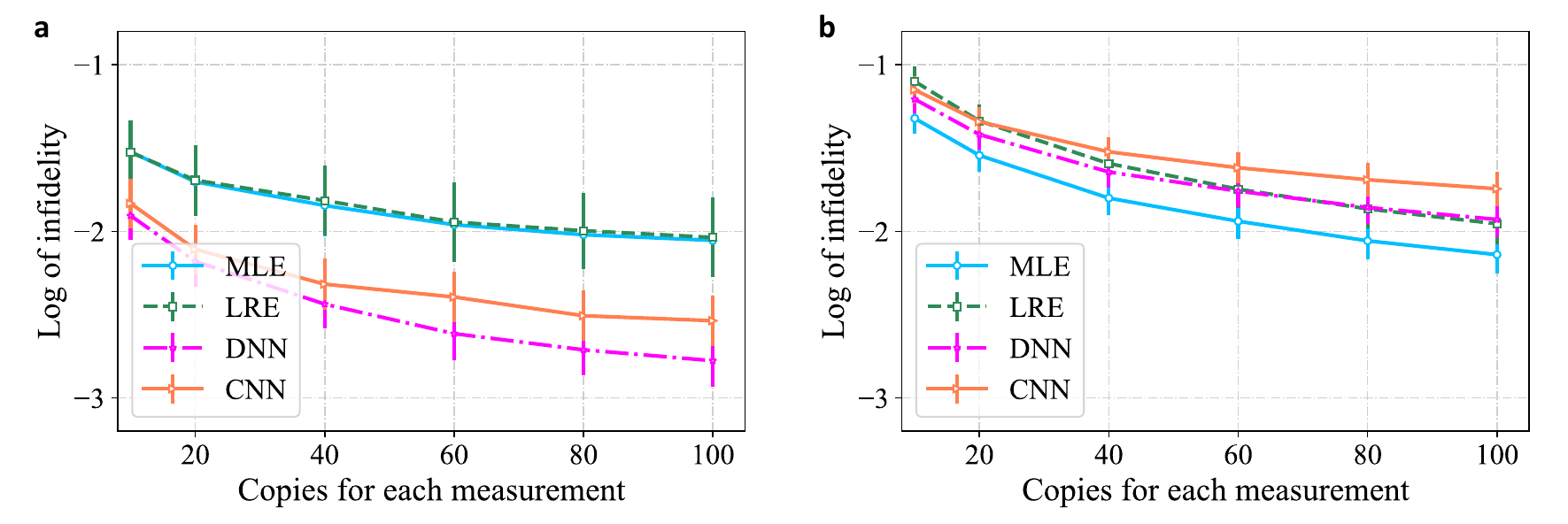}
	\caption{Comparison of several methods on 2-qubit states using cube measurements.
	\textbf{a}. Infidelity vs copies of each measurement for pure states; \textbf{b}.  Infidelity vs copies of each measurement for mixed states.}
\label{fig:case1shotscube}
\end{figure}

\begin{figure}[h]
	\centering
	\includegraphics[width=1.0\textwidth]{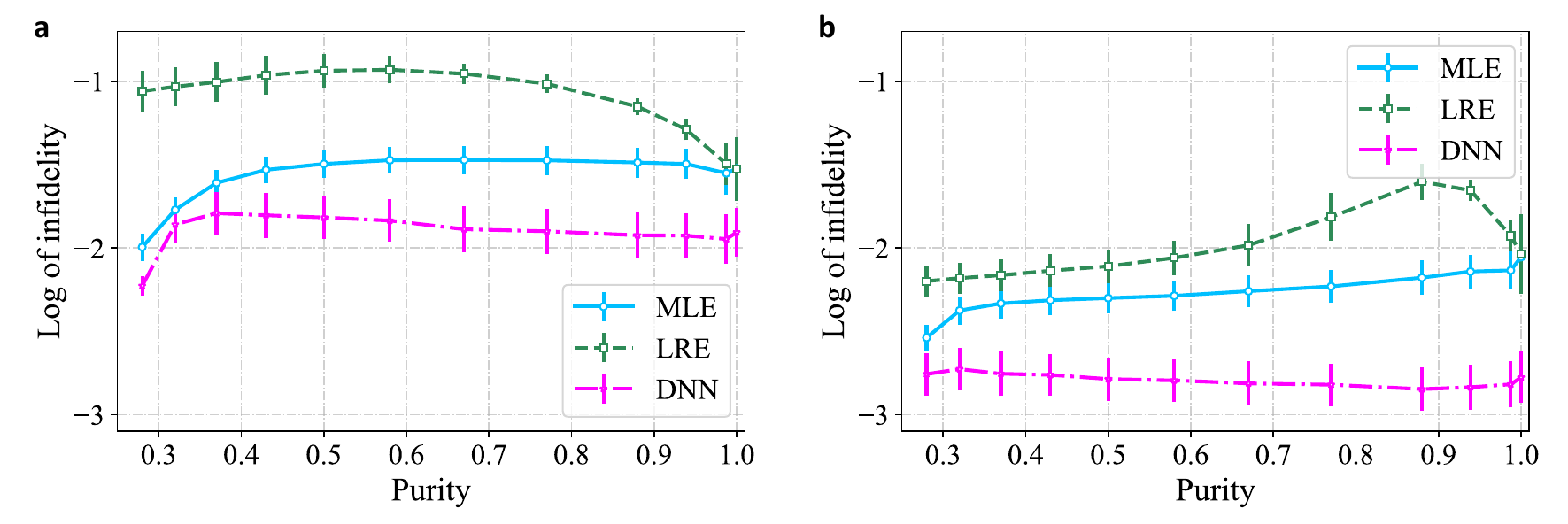}
	\caption{The performance for 2-qubit states with different purities using few measurement copies under the cube measurement. 
	\textbf{a}. Infidelity vs purity  for cube basis with $S=10$; \textbf{b}. Infidelity vs purity for cube basis with $S=100$.}
\label{fig:case1differentpuritycube}
\end{figure}

\begin{figure}[h]
	\centering
	\includegraphics[width=1.0\textwidth]{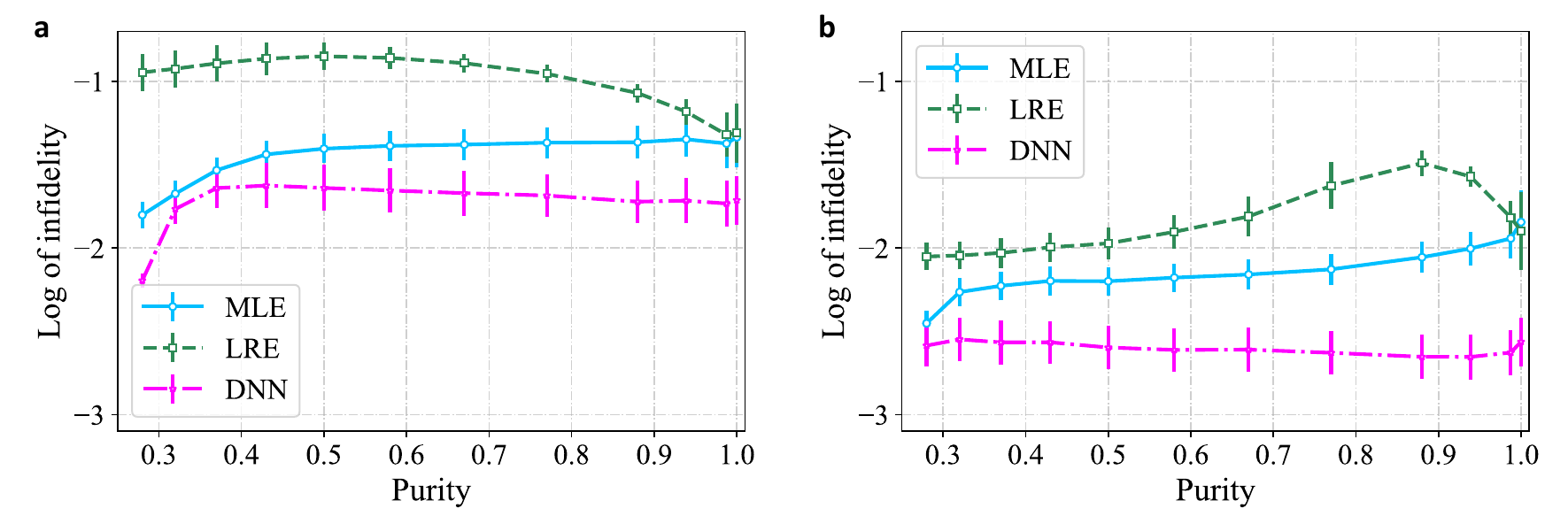}
	\caption{The performance for 2-qubit states with different purities using few measurement copies under the mub measurement. 
		\textbf{a}. Infidelity vs purity  for MUB basis with $S=10$; \textbf{b}. Infidelity vs purity for MUB basis with $S=100$.}
	\label{fig:case1differentpuritymub}
\end{figure}

\begin{figure}
	\centering
	\includegraphics[width=0.6\textwidth]{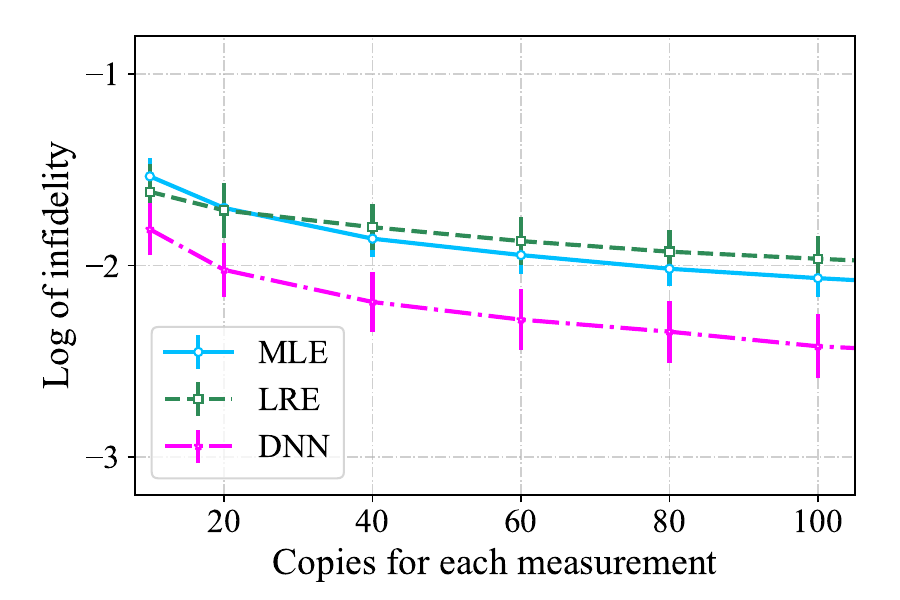}
	\caption{Results for 3-qubit random pure states when the cube measurement has few copies.}
	\label{fig:3qubitfewcopies}
\end{figure}

We first compare the performance of different methods under different measurement copies, with the results of cube measurements presented in Fig.~\ref{fig:case1shotscube}. The error bars in all of the figures represent half of the standard deviation. It is clear that DNN exhibit advantage over CNN for both random pure states and random mixed states, and the gaps increase with $S$. For pure states, our method achieves a superior performance to MLE and LRE. For mixed states, our method achieves a comparative performance to MLE. Then, we investigate the performance of mixed states with different purities. Considering the high compuation of the CNN method and the similarity of pure states and mixed states in the form of Eq.~(\ref{eq:mixed states}), we focus on the comparsion of our method against the two traditional methods.  Fig.~\ref{fig:case1differentpuritycube} and Fig.~\ref{fig:case1differentpuritymub} summarize the results of $S=10$ and $S=100$ for cube measurement and MUB measurement, respectively. DNN achieves the best performance, followed by MLE. Also, graphs of infidelity vs purity for the three methods exhibit different trends. In particular, the infidelity of MLE increases with purity, while the infidelity of LRE first increases slowly and then drops greatly with purity. By comparison, the infidelity of DNN first increases until around purity of 0.35 and drops slowly with purity for $S=10$, and the infidelity of DNN nearly keeps the same with a tiny fluctuation for $S=100$. {\color{blue}One should note that the discrepancy of results for mixed states in  Fig.~\ref{fig:case1shotscube} and Fig.~\ref{fig:case1differentpuritycube}, which arises from the different classes of mixed states, i.e., the total mixed states in Eq.~(\ref{eq:ginibre state}) and mixed states with different purities in Eq.~(\ref{eq:mixed states}).}  Fig.~\ref{fig:3qubitfewcopies} summarizes the results for QST with few copies for 3-qubit pure states following the Haar metric, where  DNN achieves the lowest log of infidelity, while MLE is better than LRE. 

\subsection{Incomplete measurements}

\begin{figure}[h]
	\centering
	\includegraphics[width=1.0\textwidth]{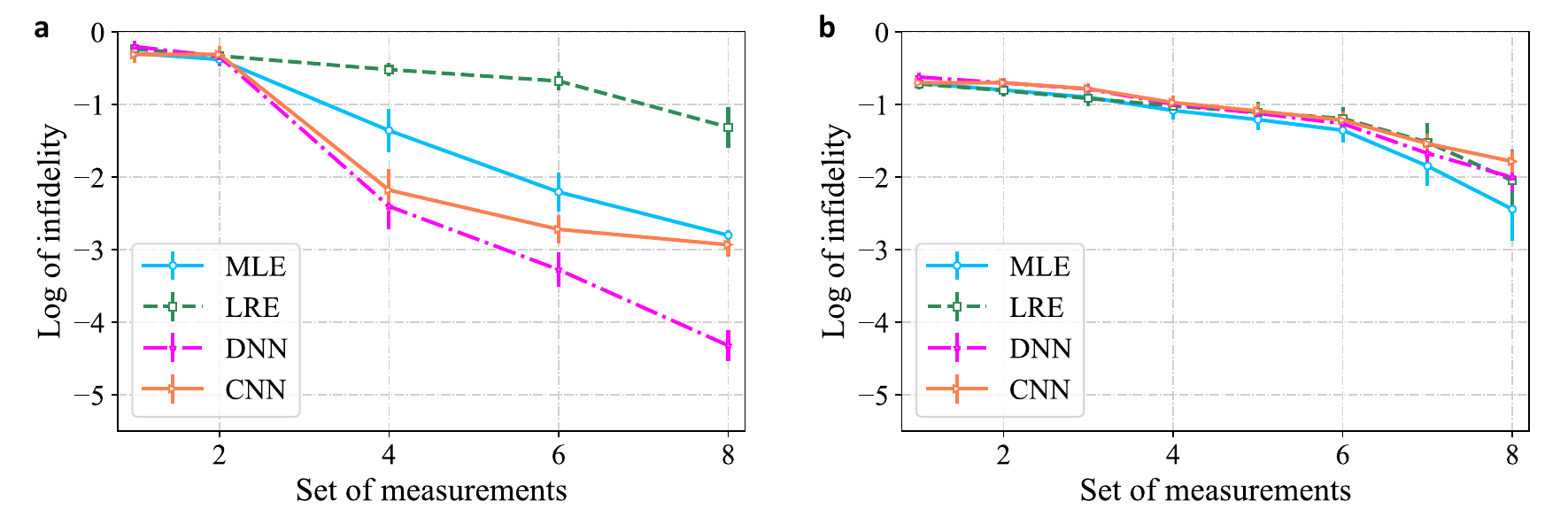}
		\caption{The performance of 2-qubit pure states and mixed states using an incomplete cube measurement. \textbf{a}. Infidelity vs sets of the cube measurement for pure states; \textbf{b}. Infidelity vs sets of the cube measurement for mixed states.}
	\label{fig:case3setcube}
\end{figure}

\begin{figure}[h]
	\centering
	\includegraphics[width=1.0\textwidth]{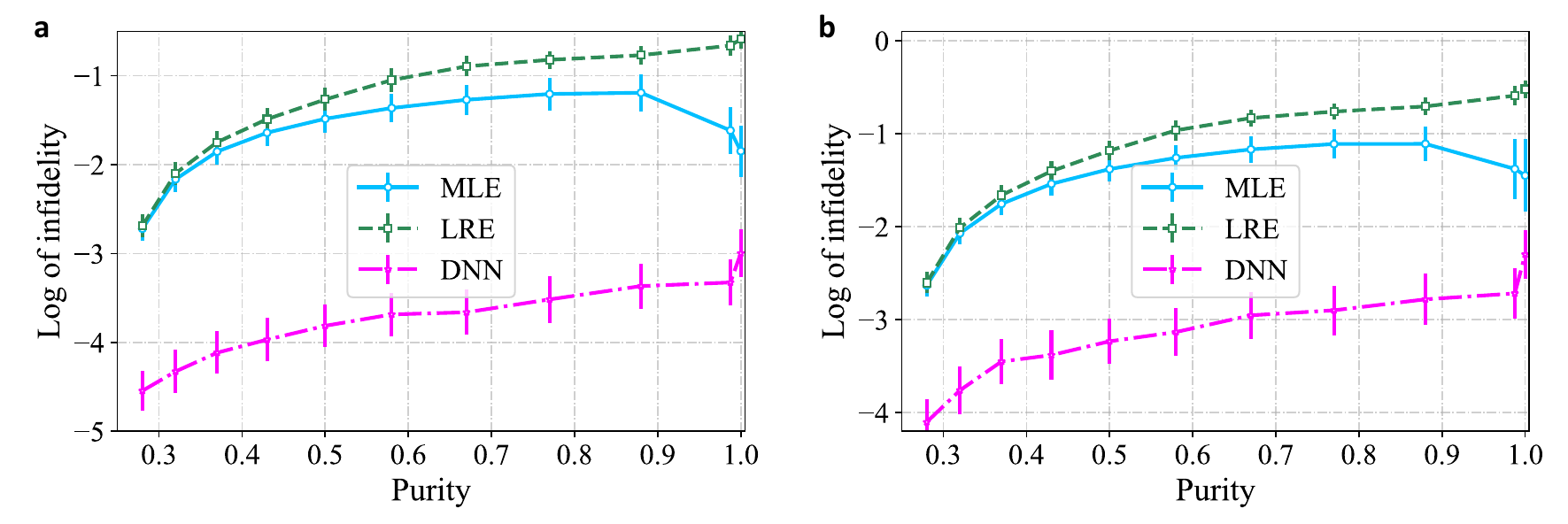}
	\caption{The performance of 2-qubit pure states and mixed states under an incomplete cube measurement.  \textbf{a}. Infidelity vs purity  for cube5; \textbf{b}.  Infidelity vs purity for mub3.}
	\label{fig:case3purity}
\end{figure}

\begin{figure}
	\centering
	\includegraphics[width=0.6\textwidth]{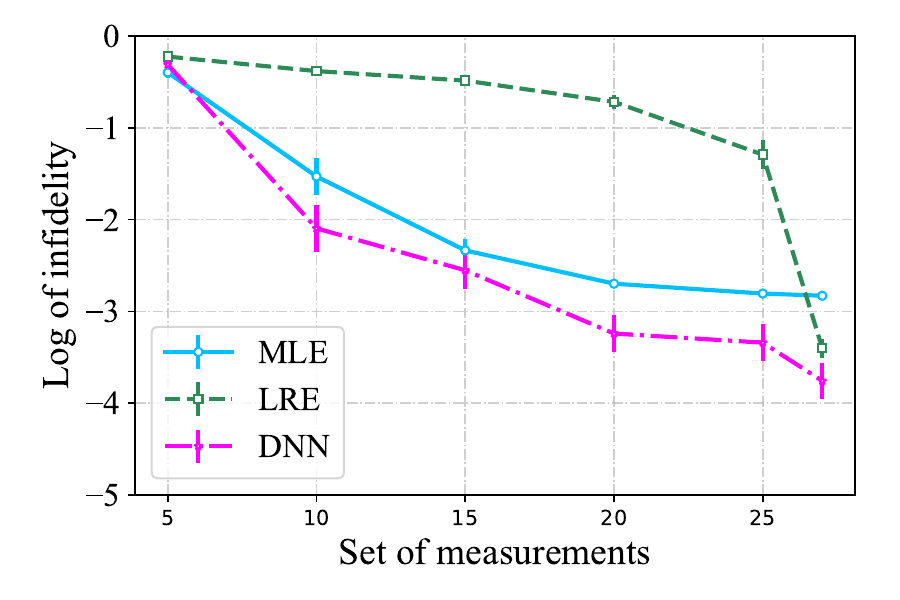}
	\caption{Results for 3-qubit random pure states using different numbers of measurement sets for the cube measurement.}
	\label{fig:3qubitincomplete}
\end{figure}

For 2-qubit systems, we consider 9 cases for the cube setting, including cube1, cube2, ..., cube9, and 5 cases for the MUB setting, including mub1, mub2, ..., mub5.  Note that cube$n$ (mub$n$) includes the first $n$ sets of projectors from the complete cube (MUB) measurement operators following the order in Tabel~\ref{Table:cube-mub}. Since the measurement projectors are usually arranged into several sets, the incompleteness of measurement operators is defined concerning the number of measurement sets.

We first investigate the comparison results of several methods based on the cube measurement.  Fig.~\ref{fig:case3setcube} demonstrates that our method exhibits better performance than CNN in addressing incomplete measurements for pure states and achieve comparable performance to traditional methods for random mixed states. Also, LRE is sensitive to the completeness of measurement operators, since it achieves the worst performance when the number of sets is 4-8 for the cube measurement.  Fig.~\ref{fig:case3purity} demonstrate that the curves of infidelity versus purity for the three methods exhibit different trends. In particular, the infidelity of DNN always increases with purity for the four cases, which is similar to that of LRE. By comparison, the infidelity of MLE increases with purity for complete measurements but exhibits a trend of first increasing and then dropping with purity for incomplete measurements.  The simulation results for 3-qubit pure states are summarized in Fig.~\ref{fig:3qubitincomplete}, where LRE fails to achieve good infidelity when the measurement set is close to the complete case, suggesting that LRE is sensitive to the completeness of measurements. 

	\section{Results on QST with noisy measurements}\label{sec:noise}


In real applications, measurements usually suffer from noise. For each measurement operator $M_i$, the existence of noise in measurements may cause it to be $\tilde{M}_i$.  For one qubit, we consider the unitary rotation operator, given by~\cite{mahler2013adaptive,chapman2016experimental,hou2020experimental}
\begin{equation}
	U(\theta_1, \theta_2, \theta_3)=\left[\begin{array}{cc}
		e^{\rm{i} \theta_1} \cos (\theta_2) & -\rm{i} e^{\rm{i} \theta_3} \sin (\theta_2) \\
		-\rm{i}e^{-\rm{i} \theta_3} \sin (\theta_2) & e^{-\rm{i} \theta_1 } \cos (\theta_2)
	\end{array}\right].
	\label{eq:rotations}
\end{equation}
For multiple qubits, we consider the local errors of unitary rotation, defined as $U_e=U_{\Theta_1} \otimes U_{\Theta_2} \cdots \otimes U_{\Theta_n}$, where $\Theta_k=(\theta_1^k,\theta_2^k,\theta_3^k)^T$ denotes the noise vector for the $k$-th qubit~\cite{lohani2020machine}.  Hence, we have the noisy measurement operator $\tilde{M}_i=U_e M_i U_e^{\dagger}$. In this case,  the measured frequencies are observed using the noisy operator $\tilde{M}_i$ rather than the ideal operator $M_i$. However, the reconstruction algorithm utilizes the measurement operator $M_i$ for reconstructing the density matrices, since the noise parameters are not available to the reconstruction algorithms.

In particular, we consider two types of unitary noise: (i) noise parameters are drawn from a Gaussian distribution (denoted as $\mathcal{N}(\mu,\sigma)$, with $\mu$ and $\sigma$ representing the mean value and the standard deviation value, respectively; (ii) noise parameters are drawn from a uniform distribution (denoted as $\mathcal{U}(\mu_1,\mu_2)$, with $\mu_1$ and $\mu_2$ representing the lower bound and the upper bound, respectively). For simplicity, a noise ratio $\xi$ is introduced to control the sampling of three ratation angles. The noise parameters for the first case are sampled as $\theta_1  \sim  \mathcal {N}(0,\pi\xi)$, 
$\theta_2 \sim \mathcal{N}(0,2\pi\xi)$, and $\theta_3\sim \mathcal{N}(0,2\pi\xi)$~\cite{lohani2020machine}. For the uniform case, the noise parameters are sampled as $\theta_1 \sim  \mathcal {U}(0,2\pi\xi)$, $\theta_2 \sim \mathcal{U}(0,0.5\pi\xi)$, and $\theta_3\sim \mathcal{U}(0,2\pi\xi)$. Here, we set $\theta_2 \in [0,0.5\pi\xi]$ owing to the symmetry of $\theta_1$ and $\theta_3$ in Eq.~(\ref{eq:rotations}).

\begin{figure}[h]
	\centering
	\includegraphics[width=1.0\textwidth]{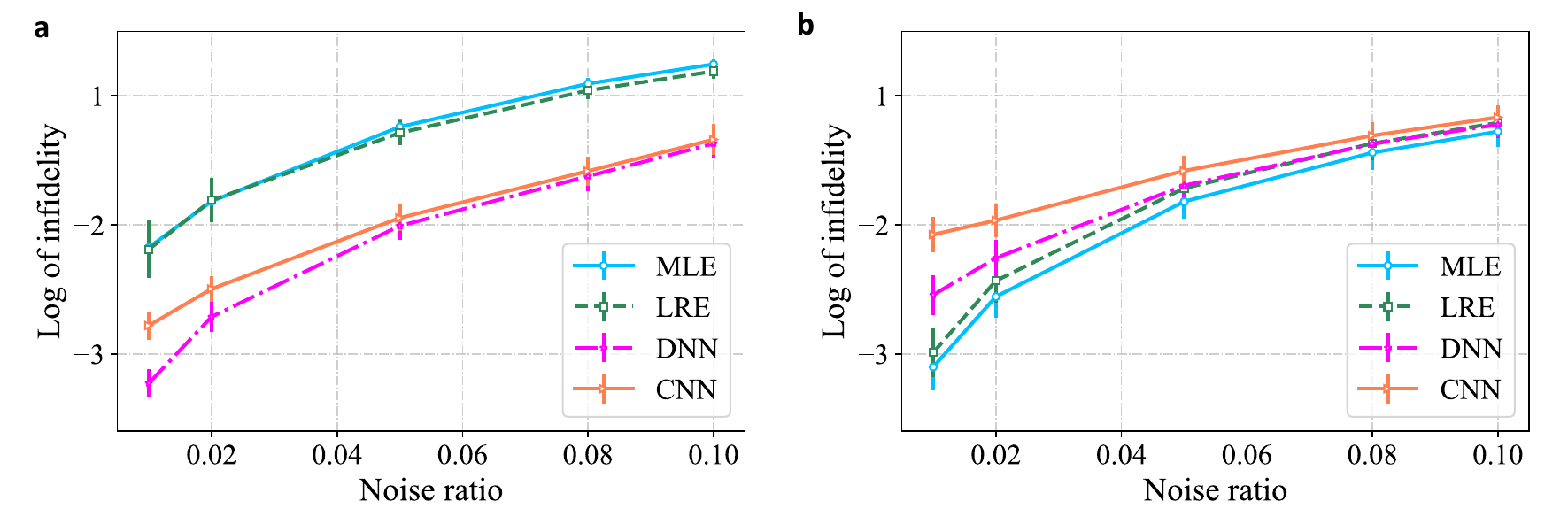}
	\caption{Comparison of several methods when the cube measurement suffers from Gaussian noise. \textbf{a}. Infidelity vs noise ratio  for pure states; \textbf{b}. Infidelity vs noise ratio for mixed states.}
\label{fig:case2noise}
\end{figure}

\begin{figure}[h]
	\centering
	\includegraphics[width=1.0\textwidth]{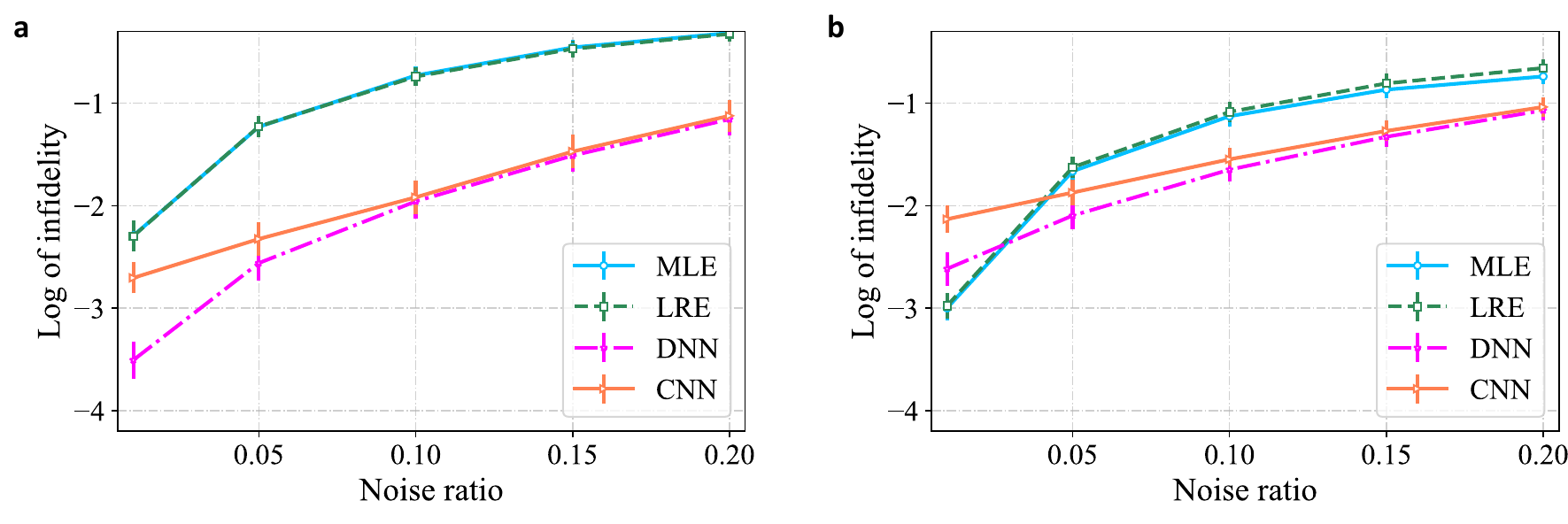}
	\caption{Comparison of several methods when the cube measurement suffers from uniform noise. \textbf{a}. Infidelity vs noise ratio for pure states; \textbf{b}. Infidelity vs noise ratio for mixed states.}
	\label{fig:case4noise}
\end{figure}

\begin{figure}[h]
	\centering
	\includegraphics[width=1.0\textwidth]{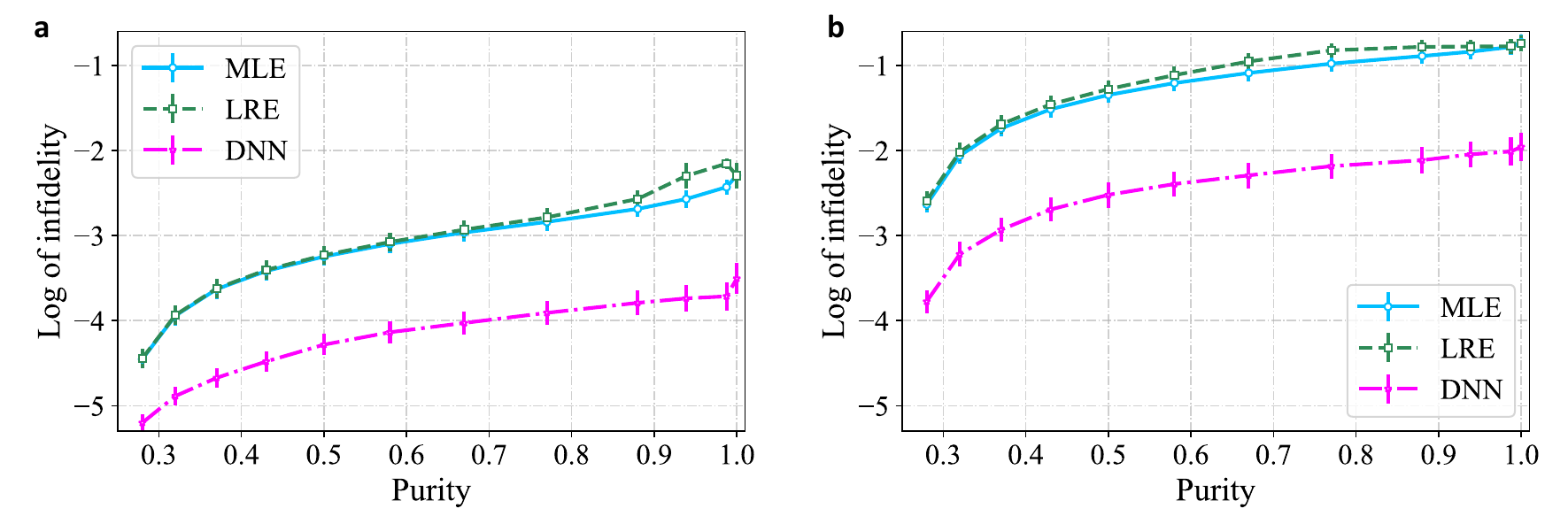}
	\caption{The performance of 2-qubit states with different purities using noisy cube measurement. \textbf{a}. Infidelity vs purity for uniform noise with ratio $0.01$; \textbf{b}. Infidelity vs purity for uniform noise with ratio $0.1$.}
	\label{fig:case4differentpurity}
\end{figure}

\begin{figure}
	\centering
	\includegraphics[width=0.6\textwidth]{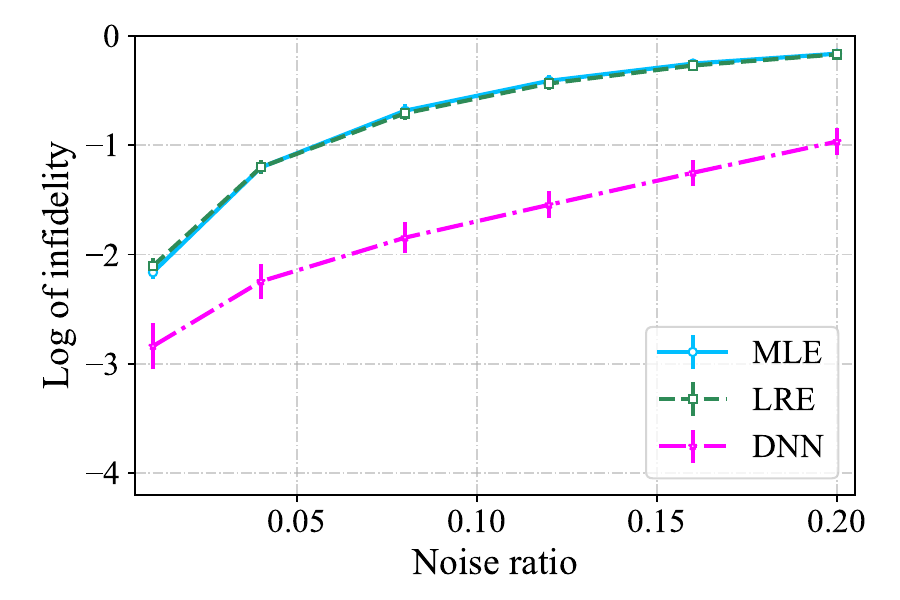}
	\caption{Results for 3-qubit random pure states when the cube measurement suffers from uniform noise.}
	\label{fig:3qubitnoise}
\end{figure}

The comparsion results when the cube measurement suffers from Gaussian noise and uniform noise are summarized in Fig.~\ref{fig:case2noise} and Fig.~\ref{fig:case4noise}, respectively.  Our method outperforms CNN for both pure and mixed states. The advantage of our method over the two traditional methods are obvious for pure states or mixed states when measurements suffer from uniform noise. Then, we focus on uniform noise and  investigate the performance of states with different purities. Fig.~\ref{fig:case4differentpurity} reveals that DNN ranks first, followed by MLE. The infidelity of DNN and MLE increases with purity, while the infidelity of LRE first increases slowly and then drops greatly with purity. The results of QST with noise for 3-qubit pure states are summarized in Fig.~\ref{fig:3qubitnoise}, where LRE and MLE achieve similar performance when the cube measurement suffers from uniform noise. Compared with the two methods, DNN exhibits a superior performance under different noise
 ratios. 

\section{Discussion and conclusion}\label{sec:conclusion}

QST is a significant task that has implications for many other quantum information processing tasks. Recognizing the challenges posed by imperfect measurement settings, we developed a general framework that leverages NNs to estimate quantum states with constrained measurements. To demonstrate its efficiency, we applied it to three cases, including a few measurement copies and incomplete measurements as well as noisy measurement operators. Numerical results demonstrate that our approach has a great potential to achieve higher efficiency than when estimating states with limited resources. Besides, our method exhibits favorable robustness when reconstructing quantum optical states with noisy measurements.  If the states being tested differ significantly from those used to train the DNN, it is anticipated that the accuracy of these testing samples will degrade. To mitigate this, transfer learning techniques can be employed to endow the NNs with improved generalization capabilities. In our future work, we will incorporate the noise distribution of the state preparation and measurement into the design of NN architecture to achieve improved estimation of quantum states under different noises. In addition, we may explore the use of ML to search for optimal measurement settings, aiming to achieve improved estimation of quantum states.

\backmatter

%
%
%

\bmhead{Acknowledgements}

We thank Zhi-Bo Hou for useful discussions.  This work was supported by the Australian Research Council's Future Fellowship funding scheme under Project FT220100656, the Australian Research Council's Discovery Projects funding scheme DP210101938, and the U.S. Office of Naval Research Global under Grant N62909-19-1-2129.

\bmhead{Data avability} The data generated in this study have been deposited in the Figshare database, which can be accessed at https://doi.org/10.6084/m9.figshare.24311704.

\section*{Declarations}

Conflict of interest No conflict of interest is declared.

\section*{Author Contributions}
H. M., D. D., and I.R. P. developed the method. C.-J. H. and G.-Y. X contributed to the part of noisy measurement. H. M. performed the simulation with the help from D. D. and C.-J. H. H. M. and D. D. drafted the manuscript with revisions from all the other authors.

\newcounter{tempfig}
\setcounter{tempfig}{\value{figure}}
%
%
%
%
%
%
%
%


%
%

\end{document}